%% file: aj_draft_24aug2011.tex
\documentclass[10pt,preprint]{aastex}

\usepackage{ulem}

\begin{document}

\title{HIGH CADENCE TIME-SERIES PHOTOMETRY OF V1647 ORIONIS}
\author{Fabienne~A.~Bastien\altaffilmark{1,2}, 
Keivan~G.~Stassun\altaffilmark{1,2},
David~A.~Weintraub\altaffilmark{1}}
\altaffiltext{1}{Vanderbilt University, Physics and Astronomy Department, Vanderbilt University, 1807 Station B, Nashville, TN 37235}
\altaffiltext{2}{Fisk University, Department of Physics, Fisk University, 1000 17th Ave. N, Nashville, TN 37208}

\begin{abstract}
We present high cadence (1--10~hr$^{-1}$) time-series photometry of the 
eruptive young variable star V1647 Orionis during its 2003--2004 
and 2008--2009 outbursts.  
The 2003 light curve was obtained mid-outburst at the phase of steepest
luminosity increase of the system, during which time the accretion rate
of the system was presumably continuing to increase toward its 
maximum rate.  The 2009 light curve was obtained after the system 
luminosity had plateaued, presumably when the rate of accretion 
had also plateaued.  
We detect a `flicker noise' signature in the power spectrum 
of the lightcurves, which may suggest that the stellar magnetosphere continued 
to interact with the accretion disk during each outburst event.  Only the 2003 
power spectrum, however, evinces a significant signal with a period of 
0.13~d.  While the 0.13~d period cannot be attributed to the stellar 
rotation period, we show that it may plausibly be due to short-lived radial oscillations 
of the star, possibly caused by the surge in the accretion rate.

\end{abstract}


\section{INTRODUCTION\label{intro}}

On 9 February 2004, \citet{mcneil04} discovered a previously
unknown object about 12 arcminutes south-west of the M78 reflection nebula.
Studies of images of the area taken prior to the event confirmed that a star,
V1647 Orionis, had brightened significantly over the course of a few days,
illuminating the material surrounding it and creating what is now known as
McNeil's Nebula.  \citet{briceno04}, through their long-term survey
of the Orion Nebula region, constrained the onset of the outburst to early
November 2003.  The lightcurve obtained by \citet{acosta-pulido07}
shows that the object reached maximum brightness by the beginning of
March 2004 and had faded back to its initial state by March 2006.
In examining photographic plates from the Asiago and Harvard Observatories,
\citet{aspin06} found that this star had undergone a similar
eruption in 1966, fading back to invisibility by the end of November 1967.
In 2009, \citet{aspin09} reported that yet another outburst event began in
August 2008.  

The rare nature of these events in general, and the fact that
we have been able to observe several of them from one star in particular,
give us a unique opportunity to study such stellar outbursts within the
context of star formation. Here, we present high cadence time-series
photometry of V1647 Orionis from its 2003--2004 and 2008--2009 eruptions.

\section{LIGHT CURVE OBSERVATIONS AND REDUCTIONS\label{observations}}

\begin{figure}[ht]
\epsscale{0.9}
\plotone{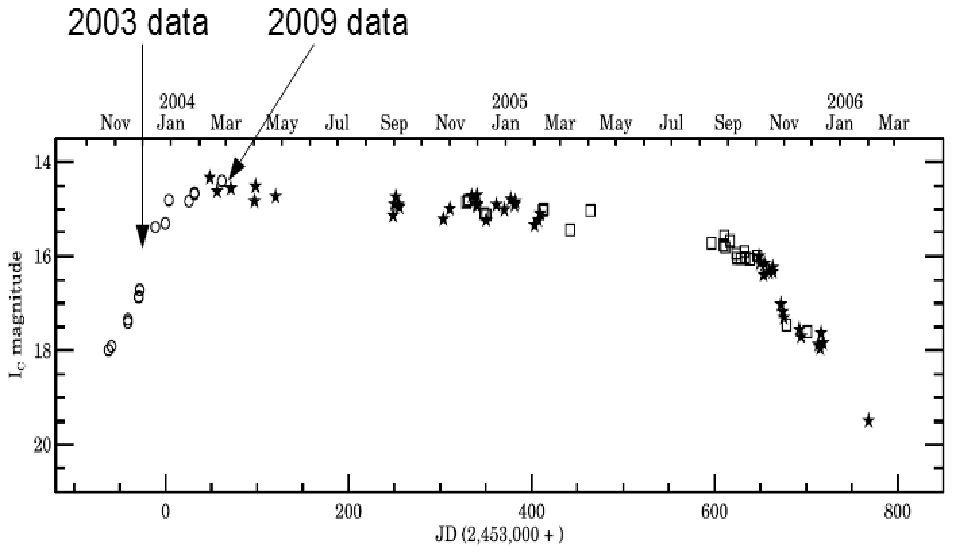}
\caption{\label{context}
Our datasets in context.  The 2003 data were taken in 2003 December, 
during the phase of steepest increase in brightness.  No outburst 
lightcurve has yet been published for the 2008-2009 event, but if we 
assume a rise-time similar to the 2003-2004 outburst, our 2009 data 
would sample the phase when the brightness of V1647 Ori had plateaued, 
as shown.  Adapted from \citet{acosta-pulido07}.
}
\end{figure}

We observed V1647~Ori during a nine-night observing run
in 2003 December with the Mosaic-1 wide-field imager
on the WIYN 0.9-m telescope at the Kitt Peak National Observatory.
This instrument consists of eight $2048 \times 4096$ pixel CCDs with a 
plate scale of 0\farcs43 pix$^{-1}$ and total field of view of
59\arcmin $\times$59\arcmin. Four of the nine nights were lost to poor
weather.  We used the SDSS $z$ filter at 9400\AA\ to obtain a total 
of 65 images on five nights with an average cadence of $\sim$1~hr$^{-1}$.  
The exposure time for all images, except those taken on the first
night, was 120~s.  On the first night, we took shorter exposures that were
60~s as well as longer ones of 180~s.  The observations were
obtained at airmasses ranging from 1.8 to 2.5.  
This dataset samples the time during which the source's brightness was
steeply increasing (cf. Fig.~\ref{context}).

We used the Y4KCam on the SMARTS 1.0-m telescope at the Cerro Tololo
Inter-American Observatory to observe V1647~Ori on UT 2009 January 23.
The $4064 \times 4064$ CCD has a plate scale of 0\farcs289 pix$^{-1}$
and a field of view of 20\arcmin $\times$20\arcmin.  We used the
$I_C$ filter for our 32 images taken with an average cadence
of $\sim$10~hr$^{-1}$.  The exposure time for each image was 300~s
and the range in airmass was 1.2 to 2.6.  If we assume a rise time
similar to the 2003 outburst, then these data were taken soon after
the object peaked in brightness (Fig.~\ref{context}).

In order to measure how the star's brightness changed with
time, we performed aperture photometry using standard IRAF
routines\footnote{IRAF is distributed by the National Optical Astronomy
Observatories, which are operated by the Association of Universities
for Research in Astronomy, Inc., under cooperative agreement with
the National Science Foundation.}.  We used an aperture radius of 6
pixels and measured the sky background with a 5-pixel wide annulus
and a 10-pixel inner radius for the 2003 data; we used an 8-pixel
inner radius for the 2009 data.  We selected these parameters based on
the average seeing of the two datasets (3.6 pixels in 2003 and 3.0 pixels
in 2009).

We performed differential photometry because the observing conditions
were non-photometric in both 2003 and 2009.  Because most of the stars 
in this field are likely variable, we did not attempt to choose a single 
comparison star with which to determine the differential light curve of 
V1647~Ori.  We expect that on average any variations in the field stars are 
uncorrelated except for effects of the instrument and sky conditions.
Thus we selected five calibration stars (Table~\ref{tab-pos})
in the vicinity of the McNeil's 
object that were of comparable brightness to V1647~Ori with which we 
defined an average ``reference star."  
We checked that none of these are known to be
variables. We further checked that the differential light curve of 
each of these stars, determined relative to the other four calibration 
stars, was not variable within the photometric errors.

The full differential light 
curve from 2003 December is presented in Table~\ref{tab-2003} and from 
2009 January in Table~\ref{tab-2009}.  The errors on individual 
photometric measurements are typically 0.02 mag in the 2003 data and 
0.01 mag in the 2009 data, which include both the formal photometric errors 
of V1647~Ori and the error of the mean of the combination of the five
comparison stars.

\section{RESULTS\label{results}}

\subsection{2003 light curve}

\subsubsection{Variability\label{variability}}

Our data show that V1647~Ori brightened by almost 2 mag over the course of 
our 9-night run in 2003 December (Fig.~\ref{fig-2003}, {\it top}). 
This steep rise in brightness is approximately linear (in magnitudes), 
but with low-level structure superposed on top of the linear trend. 
To explore the structure of these low-level brightness variations, we
de-trended the light curve as follows. 

First we subtracted a linear fit (Fig.~\ref{fig-2003}, {\it top}, dashed line), 
which reveals a slow brightness variation with a peak-to-peak amplitude 
of $\sim$0.3 mag (Fig.~\ref{fig-2003}, {\it middle}).
These variations are reminiscent of those commonly observed in classical
T Tauri stars, which can arise from variations in the accretion stream
or from modulation due to star spots at the stellar rotation period,
and which often exhibit periodic or quasi-periodic behavior 
\citep[Type IIp and Type II, respectively, in the nomenclature of][]{herbst94}
on timescales of $\sim$1--10~d.
The V1647~Ori variations appear to modulate on a timescale of 4--5~d.
We cannot establish with our data whether this signal is strictly periodic
because our light curve spans only two cycles of such a period, and 
moreover the data gaps in the light curve leave large phase gaps when
the light curve is folded on such a period.
Furthermore, we show below that this light curve modulation cannot be the
rotation period of the star. 
In the following we refer to this component of the light curve variability
as a ``quasi-periodic" modulation. We defer speculation about its possible 
physical significance to Sec.~\ref{discussion}.

Next we further de-trended the light curve by subtracting a best-fit
sinusoid as a simple representation of the quasi-periodic modulation
($P$=4.14~d; Fig.~\ref{fig-2003}, {\it middle}, dashed curve).
The resulting residual light curve (Fig.~\ref{fig-2003}, {\it bottom})
reveals very short-timescale variations 
that are significantly larger than the noise in our data
(reduced chi-square is $\chi^2_\nu = 9.7$).  
The amplitude of these 
variations is $\sigma_{\rm rms} \approx 0.05$ mag.


We conducted a few simple tests to verify that none of the periodic photometric 
signals discussed here and in what follows correlate with seeing variations.  
We performed aperture photometry on a portion of the nebula itself using, as 
before, a 6 pixel aperture radius; we separately performed aperture photometry 
on V1647 Ori using a larger (9 pixel) aperture radius.  We executed a periodogram 
analysis (as below) on the resultant light curves and also on the seeing variations, 
which were obtained by measuring the average FWHM of our calibration stars in 
each image.  We found no significant differences in the target light curve, 
and no evidence for significant periodicities in either the nebular light curve or 
in the seeing variations (and in particular not at the periods reported below).  
We conclude that changes in seeing are not driving the periodic photometric 
variations we report in this work.

\subsubsection{Periodogram analysis}

\begin{figure}[ht]
\epsscale{0.8}
\plotone{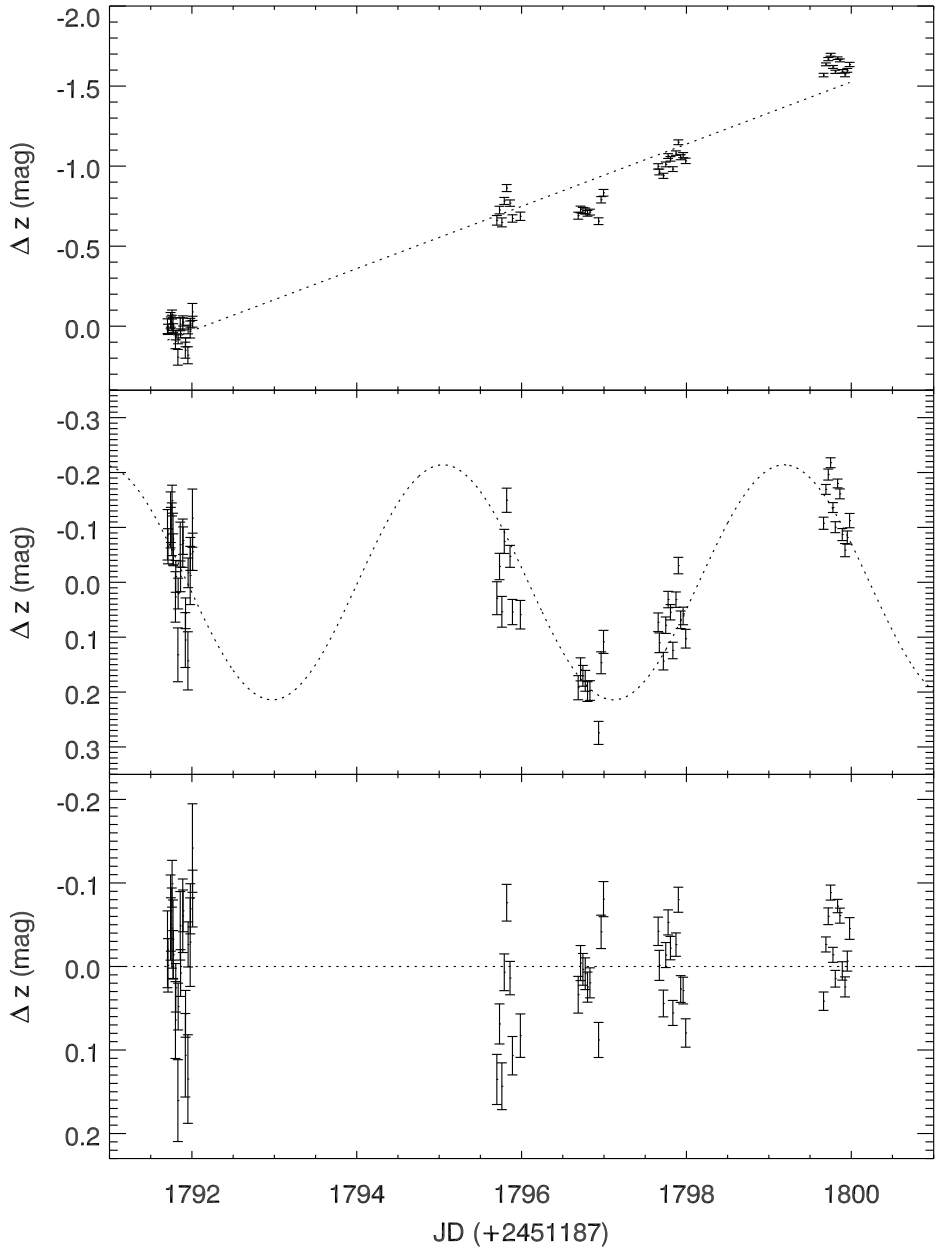}
\caption{\label{fig-2003}
{\it (Top)}: Differential light curve of V1647~Ori obtained mid-outburst
in 2003 December. The system brightened by $\sim 2$ mag over the course
of 9 nights. For reference, the dotted line represents a linear trend
fit to the data. The scale of the time axis has been chosen to allow
direct comparison with the long-term light curve data presented in
\citet[][cf.\ their Fig.~3]{briceno04}.
{\it (Middle)}: Same light curve de-trended with the linear trend from 
{\it top}, revealing a quasi-periodic modulation with a timescale of
$\sim$4~d. For reference, the dotted curve represents a sinusoidal
trend fit to the data ($P$=4.14~d). 
{\it (Bottom)}: Same light curve further de-trended with the sinusoidal
trend from {\it middle}, revealing short-timescale variations with
amplitude $\sim$0.05 mag.
}
\end{figure}

To examine the light-curve variations of V1647~Ori in detail, 
we subjected the 2003 light curve (Fig.~\ref{fig-2003})
to a standard Lomb-Scargle power spectrum analysis. 
The Lomb-Scargle periodogram is well-suited to unevenly sampled data 
such as ours. It moreover possesses well characterized statistical
properties that permit quantitative assessment of the statistical 
significance of any periodic behavior in the data
\citep[see][and references therein]{press92}.
We will exploit these statistical properties below.

\begin{figure}[ht]
\plotone{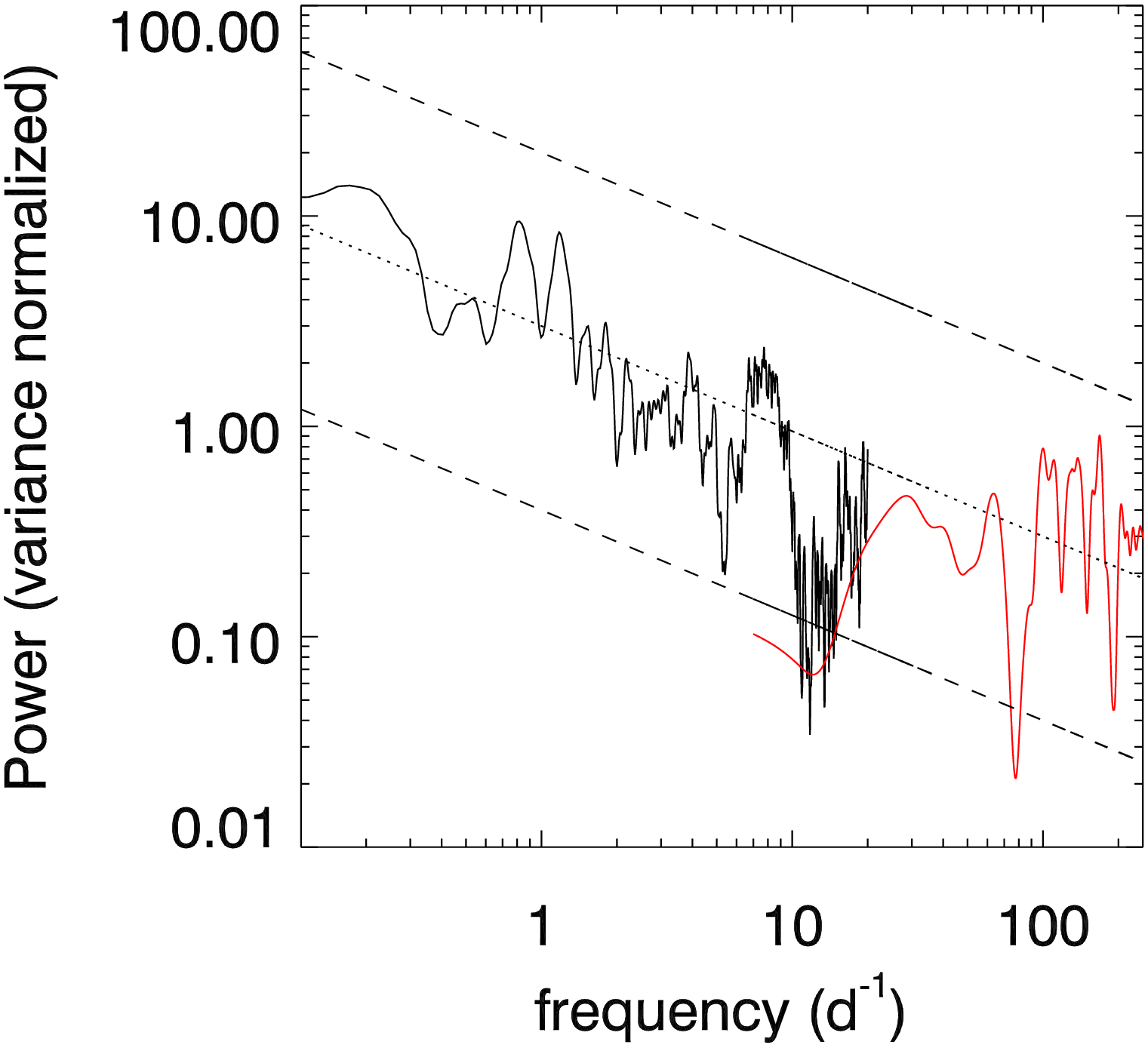}
\caption{\label{fig-flicker03}
Power spectrum resulting from a Lomb-Scargle periodogram analysis of the 
non-detrended light curve of V1647~Ori, on a log-log scale.
The frequency range $0.1 < f < 18$~d$^{-1}$ (black) is provided by the 2003 light curve
data (Fig.~\ref{fig-2003}, {\it top}), whereas the frequency range 
$7 < f < 135$~d$^{-1}$ (red) is provided by the high-cadence 2009 light curve data
(Fig.~\ref{fig-2009}).
Following the usual definition of the Lomb-Scargle periodogram, 
the ordinate gives the spectral power normalized by the variance of the data.
The power spectrum overall follows a slope of $1/\sqrt{f}$ (dashed/dotted lines),
and in addition exhibits peaks near frequencies of 
$\sim$ 0.2, 0.8, 1.2, and 8 d$^{-1}$ (see the text for discussion of
the meaning of these peaks).
}
\end{figure}

Fig.~\ref{fig-flicker03} shows the power spectrum of the 
non-detrended light curve (Fig.~\ref{fig-2003}, {\it top})
over the frequency range 0.1--18~d$^{-1}$.
The low frequency cutoff corresponds to
$1/T$ where $T$ is the total timespan of the data while the
high frequency cutoff corresponds to half the sampling
frequency (i.e.,\ the Nyquist limit).
Overall the power spectrum rises toward smaller frequencies, with a
slope that closely approximates a $1/\sqrt{f}$ dependence (represented
by dashed/dotted lines in the figure).
In addition, the power spectrum exhibits several peaks on top of the
$1/\sqrt{f}$ slope. The broad peak at $f\sim 0.2$ d$^{-1}$ corresponds 
to the slow, quasi-periodic modulation discussed in Sec.~\ref{variability}
(see also Fig.~\ref{fig-2003}, {\it middle}).
The two peaks near $f=0.8$ d$^{-1}$ and $f=1.2$ d$^{-1}$ are aliases of the
$f\sim 0.2$ d$^{-1}$ modulation beating against the diurnal data gaps 
($f=1$ d$^{-1}$).
The peak near $f=8$ d$^{-1}$ is due to the short-timescale variability
in the detrended light curve (Fig.~\ref{fig-2003} {\it bottom}). 

\begin{figure}[ht]
\epsscale{0.7}
\plotone{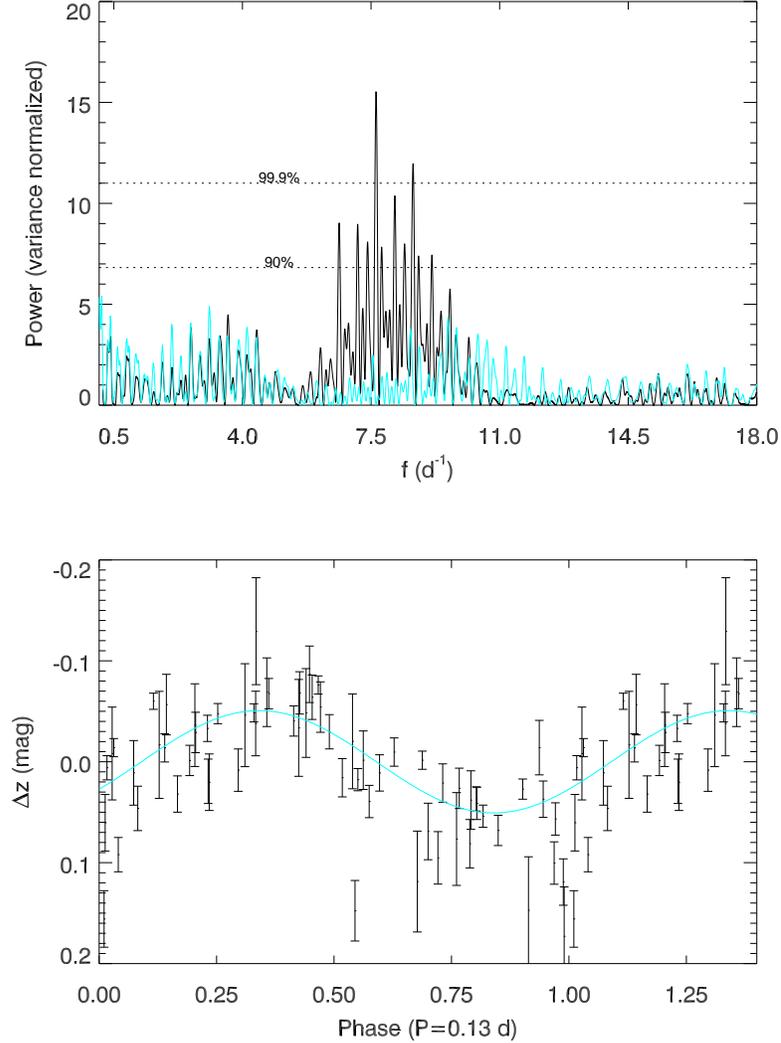}
\caption{\label{fig-period}  
{\it Top:} Power spectrum of the
detrended light curve from Fig.~\ref{fig-2003}c (black). 
The highest peak is at 7.7~d$^{-1}$ (a period of 0.13~d).  
Dotted lines represent the peak heights corresponding to
confidence levels of 90\% and 99.9\%; a peak above the 99.9\%
confidence line, for example, would have a false-alarm probability 
(FAP) lower than 0.1\%. 
The signal at 0.13~d,
is very highly statistically significant, with a FAP of $1.3 \times 10^{-5}$.
(See the text and Fig.~\ref{fapplots} for FAP details.)
The blue curve shows the result of removing the 0.13~d period; no
significant periods remain.  
{\it Bottom:} Light curve data
are phased on the 0.13~d period with best-fit sinusoid in blue.
The amplitude of the sinusoid is 0.051 mag.}
\end{figure}

The power spectrum of the detrended light curve 
is shown in Fig.~\ref{fig-period} ({\it top}, black curve).
Not surprisingly, nearly all of the power at low frequencies has been 
eliminated by the de-trending of the linear rise and of the slowly varying 
quasi-periodic modulation.  
The power spectrum shows a strong peak at 7.7~d$^{-1}$, corresponding
to a period of 0.13~d, and several other strong features at nearby
frequencies, which have power levels corresponding to a statistical
confidence of 90\% or higher (see below). 
When we filter out the 0.13-d period peak by subtracting the best
fitting sinusoid from the light curve, all of the other statistically
significant peaks in the periodogram are also removed
(Fig.~\ref{fig-period}, {\it top}, blue curve),
showing them to be aliases and beats of the 0.13~d period.  
The light curve (Fig.~\ref{fig-2003}, {\it bottom}) is
shown folded on this period in Fig.~\ref{fig-period} ({\it bottom}),
with the best-fitting sinusoid overlaid in blue. 
The amplitude of this sinusoid is 0.051~mag.

To empirically determine the false-alarm probability (FAP)
corresponding to different power levels in the periodogram, we used
a Monte Carlo bootstrapping technique as described in \citet{press92}
and implemented in e.g.\ \citet{stassun99,stassun02}.  We generated
10,000 artificial light curves by shuffling the actual measurements
in temporal order and sampling at the same timestamps as the actual
data. In this way, the artificial light curves retain both the noise
properties and the time windowing of the real data.   
For each of the 10,000
artificial light curves, we calculated a power spectrum as for the real
data and recorded the power level of the strongest peak in each. The
resulting distribution of these 10,000 maximum peak heights gives
directly the probability of a given peak height occurring by chance.

\begin{figure}[ht]
\plotone{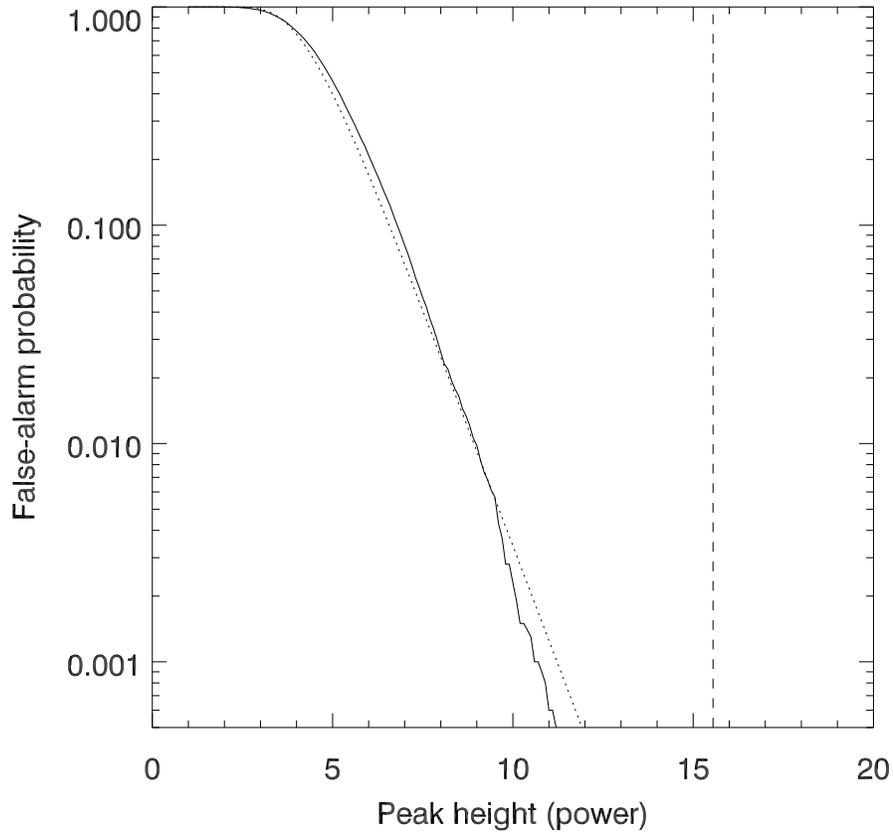}
\caption{\label{fapplots}
The false alarm probability (FAP) of the 0.13~d period as a 
function of peak height as determined from Monte Carlo simulations.  The 
solid line represents the probability determined from 
our data, and the dotted curve, which nearly overlays the solid line, 
shows the analytical probability described by \citet{press92}.  The vertical 
dashed line shows the peak height we actually observe in 
Fig.~\ref{fig-period}a.  
The FAP of the 0.13~d period is $1.3\times 10^{-5}$.
}
\end{figure}

The distribution of maximum peak heights is shown in
Fig.~\ref{fapplots} for 10,000 artificial light curves
simulating the detrended 2003 light curve (Fig.~\ref{fig-2003},
{\it bottom}).  From the figure, peak heights with power levels
above $\sim$7 occur in fewer than 10\% of the simulated power
spectra, whereas peaks with power levels above $\sim$11 occur in
fewer than 0.1\% of the simulated power spectra; these then define
the 10\% and 0.1\% FAP levels, respectively (or equivalently, the
90\% and 99.9\% confidence levels).  These confidence levels are
represented by horizontal dotted lines in the observed power spectrum
(Fig.~\ref{fig-period}, {\it top}).

As described by \citet{press92}, the FAP for a given peak height in a
Lomb-Scargle periodogram is expected to follow an analytic relationship 
(cf.\ their Eq.~13.8.7) shown in Fig.~\ref{fapplots} as a dotted curve.
Evidently, our data and its associated power spectrum closely follow 
the expected statistical behavior.  Fig.~\ref{fapplots} 
shows the FAP calculation for the 0.13~d period.  The observed peak
height (shown as vertical dashed line) has a FAP of $1.3 \times
10^{-5}$, and is therefore very highly statistically significant.

We checked that the 0.13~d period and its statistical significance 
are not dependent on the details of the sinusoidal de-trending that 
we performed in Sec.~\ref{variability}.  As a simple alternative 
to the sinusoidal detrending (see Fig.~\ref{fig-2003}, {\it middle}), 
we instead simply shifted all of the data points from a given night 
such that the mean differential $z$ magnitude for the observations 
made on each night was 0.0.  We then performed 
the same periodogram analysis as above.  We recovered the same 
0.13~d period as before with a FAP of $3.6 \times 10^{-5}$, again 
highly statistically significant.

\subsubsection{Stellar Rotation\label{rotation}}
Since periodic signals in young, low-mass stars are often associated
with rotation, we explored the possibility that we have detected stellar
rotation in our light curve.  
The rotation periods of low-mass pre--main-sequence stars are most 
typically in the range $\sim$2--10~d \citep[e.g.,][]{herbst02}, 
though some young low-mass stars have been observed to rotate with 
periods as short as $\sim 0.1$~d \citep{stassun99}.

Adopting values for the visual extinction ($19 \pm 2$~mag), 
K-band veiling ($1.5 \pm 0.2$~mag), spectral type ($M0 \pm 2$ subclasses), 
and K-band apparent magnitude (9.9 mag in February 2007) as determined 
by \citet{aspin08},
a bolometric correction of $BC_V = -1.3$ and a $(V-K)$ color of 3.7
as appropriate for an M0 spectral type \citep{kenyon95}, the equations for
absolute $K$ magnitude and bolometric luminosity from \citet{greene97},
and a distance of $426 \pm 20$~pc \citep{Menten07,Kraus07,Hirota07},
we calculate a stellar radius of $4.2^{+1.0}_{-1.3}$~R$_\odot$.

\citet{aspin09} observed line broadening in the spectrum of V1647~Ori
of 120 km~s$^{-1}$. While the spectra of EXor and FUor eruptive variables can 
be dominated by the hot ``atmosphere" of the inner accretion disk during
outburst, the \citet{aspin09} observations were obtained during quiescence
of the system; we therefore presume that the observed spectral broadening 
is stellar in origin.  This then provides a lower limit on the stellar rotational velocity of
$v \sin i$=120 km~s$^{-1}$.
We can also assume an upper limit from break-up considerations of 
$v_{\rm rot}$$\approx$190 km~s$^{-1}$.

With these upper and lower bounds on the rotational velocity of V1647~Ori, 
if we simultaneously push all other measured parameters to their 
1~$\sigma$ limits so as to produce the shortest and longest possible 
rotation periods, we find with greater than 99.9\% confidence that the 
stellar rotation period lies between 0.6 and 2.7 days.  
To test the robustness of our calculations, we also calculated the stellar 
radius using the quiescent V-band extinction and K-band apparent magnitude 
obtained by \citet{abraham04} prior to the 2004 outburst ($\sim$13 and 
10.3 magnitudes, respectively); from this, and assuming an inclination angle of 
$61^{\circ} \pm 14^{\circ}$ as found by \citet{acosta-pulido07}, we obtain a 
most likely rotation period of $\sim$1~d, consistent with the above range.  
If this is indeed the rotational period 
of the star, the diurnal gaps in our lightcurve would preclude its detection.
In any case, we conclude that the 0.13~d period, and the $\sim$4~d quasi-periodic 
modulation, do not correspond to the rotation period of V1647~Ori.

\subsection{2009 light curve}

\begin{figure}[ht]
\plotone{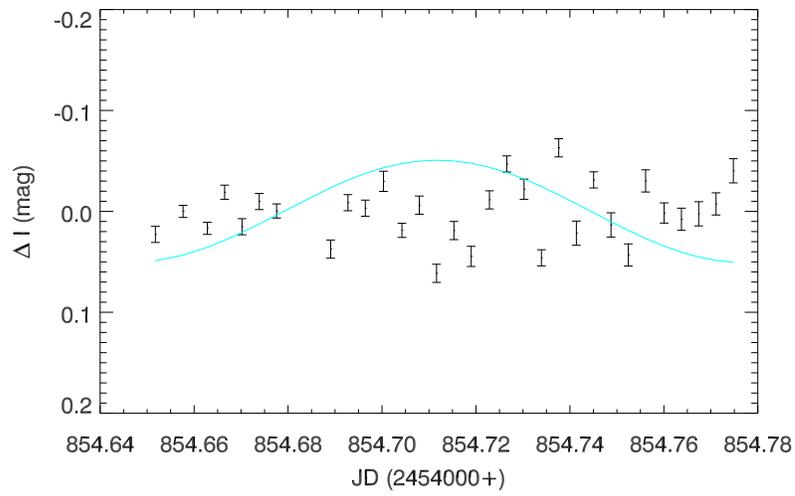}
\caption{\label{fig-2009}
Lightcurve of V1647~Ori during 2008--2009 outburst, spanning 0.13~d.
These data were taken after the object had reached maximum brightness.  
No obvious periodic brightness variations are evident.
The best-fit 0.13~d sinusoid observed in the 2003 outburst data (Fig.~\ref{fig-period}b)
is overplotted here for comparison.}
\end{figure}

Figure~\ref{fig-2009} displays our 2009 light curve. The dataset was 
obtained at very high cadence ($\sim 10$~hr$^{-1}$)
over a timespan of several hours on a single night and, thus, is sensitive 
only to high frequency variations. 
A simple $\chi^2$ test shows the light curve to be variable, with 
reduced $\chi_\nu^2 = 10.0$.
However, applying the same periodicity analysis
as above, we find no significant periodicity in this light curve 
across the frequency range to which we are sensitive, from 7~d$^{-1}$ 
to 135~d$^{-1}$; the strongest peak we obtain has a FAP of 41\%.  

The full duration of the 2009 light curve is (coincidentally) 0.13~d 
(7.7~d$^{-1}$), and thus can be used to check for the presence of a 
0.13~d period during the 2009 observations. As shown in Fig.~\ref{fig-2009},
a 0.13~d period such as that observed in the 2003 light curve
(Fig.~\ref{fig-period}b) is not present in the 2009 light curve.  
As an additional check, we injected a 0.13~d sinusoid of varying 
amplitude into these data and found that we were able to induce a 
significant peak in the power spectrum (FAP $<$ 1\%) only if the 
amplitude of the sinusoid is larger than 0.027~mag. 
We thus conclude that there is no 
evidence for an 0.13~d period in the 2009 data with an amplitude greater 
than 0.027~mag, and we can definitively rule out an 0.13~d signal with
an amplitude of 0.05 mag as seen in the 2003--2004 outburst light curve
(Fig.~\ref{fig-2003}).
To summarize, we find that this 2009 lightcurve is variable, but no periodic 
phenomenon with a frequency in the range to which we are sensitive (7~d$^{-1}$
to 135~d$^{-1}$) is driving this variability.

In Fig.~\ref{fig-flicker03} we show the power spectrum of the 2009 light
curve data together with that from the 2003 data. 
The two datasets sample a mostly disjoint range of temporal frequencies.
However, the two power spectra together are consistent with a single
$1/\sqrt{f}$ behavior for the power spectrum as a whole. 
We suggest one possible interpretation for this $1/\sqrt{f}$
behavior below (Sec.~\ref{flickering}).
The only clearly evident difference between the 2003 and 2009 power spectra
is in the overlap region near $f=10$~d$^{-1}$; as discussed above 
the 2003 light curve exhibits a strong 0.13~d period whereas the 2009 light
curve does not.



\section{DISCUSSION\label{discussion}}

We have found strong evidence to suggest that V1647~Ori exhibited
a highly significant 0.13~d periodicity in brightness
during the rapid brightening phase of its 2003 outburst.  
The amplitude of this periodic variation was $\sim$0.05 mag.
The presence of this feature in the 2003 data, obtained while 
the object was in the brightening phase, and its absence in the 
2009 data, obtained after V1647~Ori had peaked in brightness, 
suggest that this phenomenon, whatever its cause, is associated 
with the unstable period of time when the brightness of V1647~Ori 
was most rapidly increasing.  
We have also found evidence for 
a $1/\sqrt{f}$ slope in the power spectrum of V1647~Ori over a large
range of temporal frequencies, $0.1 < f < 135$~d$^{-1}$.


In this section, we consider whether the 0.13~d period may be
ascribed to short-lived stellar pulsations, perhaps triggered by the
high accretion event.  Next, we discuss 
``flickering" in the light curve, 
evidenced by the $1/\sqrt{f}$ slope in the power spectrum,
in the context of a magnetically channeled accretion flow. Finally,
we consider whether oscillations originating in the accretion disk,
similar to those observed during cataclysmic variable (CV) star outbursts,
could be observed in young star outbursts, including FUor and EXor events,
and we speculate that the $\sim$4~d
quasi-periodic modulation observed in the 2003 light curve
could correspond to such an oscillation.
The consideration of these mechanisms here is speculative; our aim is to
examine whether these explanations may be plausible, but we cannot yet establish
that these are definitive driving mechanisms for the observed variability.

\subsection{Stellar pulsations\label{pulsation}}
Since it is inconsistent with the likely rotation period of V1647~Ori
(Sec.~\ref{rotation}),
we explore the possibility that the 0.13~d period might be a
manifestation of pulsation.  Given its
effective temperature and estimated mass of $0.8 \pm 0.2$ M$_{\odot}$
\citep{aspin08}, and based on comparison with stellar evolutionary
tracks \citep{dantona94}, we find that V1647~Ori may lie just
within the theoretically predicted deuterium instability strip.  
The expected fundamental mode
pulsation period would be approximately 0.5~d \citep[e.g.,][]{Toma72},
somewhat longer than the period we detect.  In addition, the 0.13~d period  
does not appear in the 2009 data. 
It is extremely unlikely that V1647~Ori transitioned from being within
the instability strip to being outside the instability strip between 
our 2003 and 2009 observations.


We next consider whether the dramatic increase in the accretion
rate could have induced short term radial oscillations of
the stellar surface. From the idealized homogeneous compressible model
presented by \citet{cox80}, the period of oscillation, for 
purely radial pulsation, varies according to  
\begin{equation}
\frac{4\pi^{2}R^{3}}{P^{2}GM} = -4 + \Gamma_{1}(2n^{2}+5n+3)
\end{equation}
where $R$ is the star's radius, $P$ is its pulsational period, $G$ is the 
gravitational constant, $M$ is the stellar mass, $\Gamma_{1}$ 
the adiabatic exponent, and $n$ the pulsational mode.  
For our purposes here, this idealized model is not significantly different 
from more sophisticated models \citep{tassoul68}.
Adopting the stellar parameters of 
V1647~Ori (see Sec.~\ref{rotation}) and assuming as one extreme assumption
that no part of the star is ionized 
($\Gamma_{1}=\frac{5}{3}$), we 
find that the 0.13~d period could correspond to pulsational modes ranging from 
2.8 to 4.7.  In order to determine the effect of ionization zones on our results, we 
looked at the extreme case of a fully ionized gas ($\Gamma_{1}=\frac{4}{3}$); in this 
case we find pulsational modes between 3.3 and 5.4.  
Hence, any errors in 
not taking ionization zones into account do not significantly change our 
results.  We therefore infer that the 0.13~d period most likely corresponds 
to a radial pulsation in the $4^{th}$ oscillation mode, but the $3^{rd}$ 
and $5^{th}$ modes are also permitted within the observational 
uncertainties in the stellar properties of V1647~Ori.

Pulsation solely in a higher overtone radial
mode, although rare, is strongly dependent on the location of the
driving mechanism \citep{kurtz06}. Examples of similar short-term
oscillation-producing phenomena include recent solar observations
\citep{Karoff08} that revealed that energetic magnetic reconnection
and coronal mass ejection events can induce short-lived high-frequency
oscillations of the solar surface. In the Sun, these are confined to
the vicinity of the triggering event \citep[e.g.,][]{Kienreich09}.
The triggering of radial oscillations by the sudden onset of 
highly energetic accretion on V1647~Ori thus provides one possible 
explanation for the observed variability.


\subsection{Flickering\label{flickering}}
Flickering is defined as random, small amplitude brightness
variations recurring on dynamical timescales \citep{kenyon00}.
Sometimes interpreted as an observable consequence of an inhomogeneous
accretion flow, flickering has been observed in CV stars and, as it
is associated with accretion, could be observable during 
similar outburst events in young stars.
Indeed, \citet{kenyon00} and \citet{rucinski08} observed
flickering in the lightcurves of FU~Ori and TW~Hya, respectively,
the latter 
through observation of a $1/\sqrt{f}$ slope in the power spectrum.
The combination of a high observing cadence and a relatively long
time baseline means that our data are sensitive to a large range of
frequencies, and, as such, any signs of this phenomenon should be
readily apparent in an analysis of the power spectrum of our data.

As discussed in Sec.~\ref{results},
the overall power spectrum of the observed phases of the 
2003--2004 and 2008--2009 outbursts of V1647 Ori follow a
$1/\sqrt{f}$ trend
(Fig.~\ref{fig-flicker03}), 
which suggests that flickering is one possible origin of 
the random variability components of the light curve. 

The nature of such ``flicker noise"
is not entirely understood; however, it appears to be linked to
situations in which a flow (in this case the accretion of material
from disk to star) is funneled or is in some way forced to pass
through a physically confined region \citep{press78}.  
Indeed, accretion in young low-mass stars is typically envisioned to occur via
magnetospheric funneling of disk material along stellar field lines that
thread the inner accretion disk \citep[e.g.][]{shu94}.
Also, the flickering observed in the CV system T~CrB, for example, has been
attributed to turbulence in the inner regions of the accretion disk
\citep{dobrotka10}.  Thus the observation of flickering in 
V1647~Ori might suggest an interaction between the stellar 
magnetosphere and the inner regions of the accretion disk during 
outburst, with material continuing to accrete onto the star along 
stellar magnetic field lines \citep[e.g.][]{shu94}.



\subsection{Dwarf Nova-Like Oscillations in the Keplerian Inner
Accretion Disk?\label{dno}}  

We have observed a modulation with a timescale of $\sim$4~d
in the 2003 light curve of V1647~Ori (Fig.~\ref{fig-2003}).
As discussed, our data do not permit us to determine whether this 
modulation is strictly periodic, or indeed whether it persists for more
than $\sim$2 cycles.  It is nonetheless a potentially
interesting quasi-period that could arise in a number of different ways.

Dwarf-nova oscillations (DNOs) are quasi-periodic brightness variations
typically observed in cataclysmic variable star outbursts.
These phenomena are associated with accretion, and they may persist
during quiescence \citep{pretorius06}.  The oscillations are not
observed in all CV outbursts, and the oscillation frequencies may
change during a high-accretion event.  Because of their relatively
short periods, DNOs are believed to originate at the inner edges
of Keplerian accretion disks.  They have only rarely been observed
during the rise of a CV outburst \citep{warner08}, but their behavior
is thought to be well described by a low-inertia magnetic accretor
model in which accretion induces variations in the angular velocity
of an equatorial accretion belt \citep{warner02}.  In addition,
so-called quasi-periodic oscillations (QPOs), which are longer
term, less coherent oscillations, have also been observed during
some CV outbursts.  There is no generally accepted model for what
causes the QPO brightness modulations, but they are thought to be
accretion disk phenomena \citep[e.g.,][]{woudt02}. In any event,
an empirical relationship links DNOs to QPOs such that $P_{\rm QPO}
\approx 15 \times P_{\rm DNO}$ \citep{warner02}.

The low-inertia magnetic accretor model predicts that the DNO
oscillation quasi-period ($P_{\rm DNO}$) increases as the accretion
rate ($\dot{M}$) decreases.  It also predicts that $P_{\rm DNO}$
corresponds to the Keplerian orbital period of the inner edge
of the accretion disk.  If we assume a $P_{\rm DNO}$--$\dot{M}$
relation such as that found empirically for the dwarf nova SS Cyg
by \citet{mauche97} and use the $\dot{M}$ found for V1647~Ori by
\citet{muzerolle05}, we find $P_{\rm DNO} \sim 5.7$~d.  We note that
the value of $\dot{M}$ might differ for our data, bringing $P_{\rm
DNO}$ closer to our observed quasi-period of 4.14~d.  For example,
using the $P_{\rm DNO}$--$\dot{M}$ relation, we calculate that $\dot{M}
\sim 10^{-4}$ M$_{\odot}$~yr$^{-1}$ would yield a period of 4.2~d.
In any event, if this model applies to the V1647~Ori system, then,
taking the mass of the central star to be $\sim 0.8~$M$_{\odot}$,
we calculate that the inner edge of the accretion disk surrounding
V1647~Ori was located $\sim$2.5 stellar radii from the star prior to
the peak of the outburst. This model could thus suggest that the 
stellar magnetosphere was still capable of holding off the inner disk, 
at least during the early stages of the 2003--2004 outburst.

Additionally, using the relation of $P_{\rm QPO} \approx 15 \times
P_{\rm DNO}$, we would expect to observe a QPO with a period
of $\sim$60~d, which is very close to the 56~d period found by
\citet{acosta-pulido07} who attributed it to dense circumstellar
clumps orbiting the star.  However, given the non-uniform sampling of
our data and the fact that these oscillations are observed over only
$\sim$2 cycles, the reality of this period, and the link between
these periodicities and those observed in white dwarf stars, is unclear.
Nevertheless, they are potentially intriguing.  Observations probing
similar timescales during FUor/EXor outbursts would allow us to further
investigate the presence of such phenomena in young star outbursts.

\section{SUMMARY\label{summary}}
In this work, we present high cadence time-series photometry of V1647
Orionis during its two most recent outburst events.  
The overall power spectrum of our 2003 and 2009 
datasets displays a $1/\sqrt{f}$ `flicker noise' spectrum.  
The detection of `flicker noise' in the power spectra of our 
light curves suggests that accretion continues to be mediated 
by the stellar magnetosphere during the observed phases of the 
2003--2004 and 2008--2009 outbursts.  This picture is bolstered by the observation
of a quasi-periodic modulation in the 2003 light curve with a timescale
of $\sim$4~d, perhaps arising from a CV-like quasi-periodic oscillation
of the inner accretion disk edge at a height of 2--3 stellar radii 
from the stellar surface.

Our Fourier analysis of our 2003 detrended lightcurve, obtained 
mid-outburst, yields a periodic variation on a timescale of 0.13~d 
that persists in spite of the dramatic rise in brightness caused by 
the outburst.  This period, detected at very high statistical 
significance, is not attributable to the expected rotation period 
calculated from other measured properties of the star.  This 0.13~d 
period is absent from our 2009 light curve.

The 0.13~d period is very coherent
in the 2003 dataset, and it is therefore likely to have been a truly
periodic phenomenon during those observations.  Given that we do not
detect this period in our 2009 light curve, obtained post-outburst, we
conclude that it is probably an accretion-induced process associated
with the epoch when the brightness of the object is increasing
(and so perhaps when the accretion rate is increasing), one
likely candidate being short-lived radial pulsations of the star.

\acknowledgments
We thank the anonymous referee, whose constructive review significantly 
improved the manuscript.  We also thank Andreas Berlind, David Furbish, 
Kelly Holley-Bockelmann, M.\ Coleman Miller, and C.\ Robert O'Dell for helpful
discussions. F.A.B.\ is supported by NSF grant AST-0849736.
K.G.S.\ acknowledges support from NSF Career grant AST-0340975 and a
Cottrell Scholar award from the Research Corporation.

\clearpage

\input{tab1.tex}

\input{tab2.tex}

\input{tab3.tex}

\end{document}

%% file: tab1.tex
\begin{deluxetable}{rr}
\tablewidth{0pt}
\tabletypesize{\scriptsize}
\tablecolumns{2}
\tablecaption{\label{tab-pos}
J2000 coordinates of stars used for differential photometry}
\tablehead{ \colhead{RA\tablenotemark{a}} & \colhead{Dec\tablenotemark{b}} }
\startdata
05:46:22  &  -00:03:37  \\
05:46:29  &  -00:03:47  \\
05:46:31  &  -00:04:21  \\
05:46:28  &  -00:09:59  \\
05:46:26  &  -00:10:25  \\
\enddata
\tablenotetext{a}{in hh:mm:ss}
\tablenotetext{b}{in dd:mm:ss}
\end{deluxetable}

%% file: tab2.tex
\begin{deluxetable}{rrr}
\tablewidth{0pt}
\tabletypesize{\scriptsize}
\tablecolumns{3}
\tablecaption{\label{tab-2003}
Differential photometric measurements of V1647~Ori in 2003 December}
\tablehead{ \colhead{JD\tablenotemark{a}} & \colhead{$\Delta z$} & \colhead{$\sigma_{\rm mag}$} }
\startdata
2978.7040  &  0.000  &  0.046  \\
2978.7100  &  0.020  &  0.032  \\
2978.7410  &  -0.034  &  0.051  \\
2978.7440  &  -0.026  &  0.032  \\
2978.7560  &  -0.020  &  0.048  \\
2978.7590  &  -0.073  &  0.028  \\
2978.7710  &  -0.005  &  0.047  \\
2978.7740  &  0.014  &  0.029  \\
2978.8000  &  0.095  &  0.046  \\
2978.8040  &  0.056  &  0.029  \\
2978.8300  &  0.194  &  0.049  \\
2978.8330  &  0.082  &  0.028  \\
2978.8580  &  -0.012  &  0.041  \\
2978.8620  &  0.045  &  0.028  \\
2978.8890  &  -0.021  &  0.044  \\
2978.8930  &  -0.026  &  0.025  \\
2978.9200  &  0.150  &  0.050  \\
2978.9230  &  0.101  &  0.028  \\
2978.9510  &  0.182  &  0.053  \\
2978.9540  &  0.021  &  0.027  \\
2978.9790  &  0.021  &  0.053  \\
2978.9810  &  -0.019  &  0.030  \\
2979.0060  &  -0.089  &  0.053  \\
2979.0090  &  -0.028  &  0.034  \\
2982.7000  &  -0.661  &  0.030  \\
2982.7320  &  -0.725  &  0.024  \\
2982.7610  &  -0.648  &  0.028  \\
2982.7900  &  -0.782  &  0.022  \\
2982.8190  &  -0.863  &  0.022  \\
2982.8600  &  -0.769  &  0.020  \\
2982.8890  &  -0.673  &  0.023  \\
2982.9850  &  -0.687  &  0.026  \\
2983.6870  &  -0.690  &  0.022  \\
2983.7150  &  -0.729  &  0.021  \\
2983.7440  &  -0.723  &  0.019  \\
2983.7720  &  -0.719  &  0.019  \\
2983.8000  &  -0.705  &  0.018  \\
2983.8290  &  -0.713  &  0.018  \\
2983.9340  &  -0.656  &  0.021  \\
2983.9650  &  -0.790  &  0.020  \\
2983.9940  &  -0.833  &  0.021  \\
2984.6570  &  -0.998  &  0.017  \\
2984.6710  &  -0.963  &  0.018  \\
2984.7200  &  -0.939  &  0.016  \\
2984.7490  &  -1.011  &  0.015  \\
2984.7780  &  -1.063  &  0.015  \\
2984.8060  &  -1.045  &  0.014  \\
2984.8350  &  -0.981  &  0.015  \\
2984.8730  &  -1.081  &  0.014  \\
2984.9020  &  -1.149  &  0.015  \\
2984.9300  &  -1.056  &  0.016  \\
2984.9600  &  -1.068  &  0.016  \\
2984.9910  &  -1.033  &  0.017  \\
2986.6640  &  -1.569  &  0.011  \\
2986.6920  &  -1.636  &  0.009  \\
2986.7210  &  -1.668  &  0.010  \\
2986.7490  &  -1.695  &  0.009  \\
2986.7780  &  -1.619  &  0.009  \\
2986.8060  &  -1.589  &  0.010  \\
2986.8340  &  -1.674  &  0.008  \\
2986.8620  &  -1.661  &  0.009  \\
2986.8910  &  -1.592  &  0.011  \\
2986.9240  &  -1.570  &  0.012  \\
2986.9520  &  -1.599  &  0.012  \\
2986.9800  &  -1.635  &  0.013  \\
\enddata
\tablenotetext{a}{Julian Date $(+2450000)$.}
\end{deluxetable}

%% file: tab3.tex
\begin{deluxetable}{rrr}
\tablewidth{0pt}
\tabletypesize{\scriptsize}
\tablecolumns{3}
\tablecaption{\label{tab-2009}
Differential photometric measurements of V1647~Ori in 2009 January}
\tablehead{ \colhead{JD\tablenotemark{a}} & \colhead{$\Delta I$} & \colhead{$\sigma_{\rm mag}$} }
\startdata
4854.6518  &  0.023  &  0.008  \\
4854.6576  &  0.000  &  0.006  \\
4854.6628  &  0.017  &  0.006  \\
4854.6665  &  -0.019  &  0.007  \\
4854.6702  &  0.015  &  0.008  \\
4854.6739  &  -0.010  &  0.008  \\
4854.6776  &  -0.000  &  0.007  \\
4854.6854  &  0.000  &  0.009  \\
4854.6891  &  0.037  &  0.009  \\
4854.6927  &  -0.009  &  0.008  \\
4854.6964  &  -0.003  &  0.008  \\
4854.7003  &  -0.030  &  0.010  \\
4854.7042  &  0.019  &  0.007  \\
4854.7079  &  -0.006  &  0.009  \\
4854.7116  &  0.061  &  0.009  \\
4854.7153  &  0.019  &  0.009  \\
4854.7190  &  0.045  &  0.010  \\
4854.7229  &  -0.011  &  0.009  \\
4854.7266  &  -0.047  &  0.008  \\
4854.7302  &  -0.022  &  0.010  \\
4854.7339  &  0.046  &  0.008  \\
4854.7376  &  -0.063  &  0.009  \\
4854.7414  &  0.022  &  0.012  \\
4854.7451  &  -0.031  &  0.008  \\
4854.7487  &  0.013  &  0.012  \\
4854.7524  &  0.043  &  0.011  \\
4854.7561  &  -0.030  &  0.011  \\
4854.7601  &  0.002  &  0.010  \\
4854.7637  &  0.008  &  0.011  \\
4854.7674  &  0.002  &  0.012  \\
4854.7711  &  -0.007  &  0.011  \\
4854.7748  &  -0.040  &  0.012  \\
\enddata
\tablenotetext{a}{Julian Date $(+2450000)$.}
\end{deluxetable}